\documentstyle[preprint,eqsecnum,aps]{revtex}

\begin{document} \draft

\title{ Low-Energy vibrational density of states of plasticized poly(methyl
methacrylate) }
\author{L. Saviot, E. Duval }
\address{LPCML, Universit\'{e} Lyon~1,  UMR-CNRS~5620  43, boulevard du 11
Novembre 69622 Villeurbanne Cedex, France}
\author{J. F. Jal}
\address{DPM, Universit\'{e} Lyon~1, UMR-CNRS~5586 43, boulevard du 11
Novembre 69622 Villeurbanne Cedex, France}
\author{A. J. Dianoux}
\address{Institut Laue-Langevin, 38042 Grenoble Cedex, France}
\author{V. A. Bershtein}
\address{Ioffe Physico-technical Institute, Russian Academy of Sciences,
St. Petersburg 194021, Russia}
\author{L. David}
\address{GEMPPM, UMR-CNRS 5510, INSA-Lyon, 69621 Villeurbanne Cedex, France}
\author{S. Etienne}
\address{LPM, UMR-CNRS 7556, Ecole des Mines, 54042 Nancy Cedex, and EEIGM,
54010 Nancy Cedex, France}
\date{\today}
\maketitle

\newpage

\begin{abstract}
The low-energy vibrational density of states (VDOS) of hydrogenated or
deuterated poly(methyl methacrylate) (PMMA) plasticized by dibutyl phtalate
(DBP) is determined by inelastic neutron scattering. From experiment, it is
equal to the sum of the ones of the PMMA and DBP components. However, a
partition of the total low-energy VDOS among PMMA	 and DBP was
observed. Contrary to Raman scattering, neutron scattering does not show
enhancement of the boson peak due to plasticization.

\end{abstract}
\newpage

\section{INTRODUCTION}

Physical studies of plasticized polymers are of interest for attaining
optimal properties of the material and revealing the nanostructure, i.e.,
the distribution of plasticizer molecules in the polymeric glass. Recently
an experimental study was carried out on the poly(methyl methacrylate)
(PMMA) plasticized with  dibutyl phtalate (DBP) \cite{Duv98}. Two
interesting results were obtained. By small angle X-ray scattering (SAXS) a
correlation peak was observed at about 1.5 $nm^{-1}$ in the plasticized
sample, while no such peak was detected in the non-plasticized one. This
observation shows that the plasticizer is not distributed homogeneously in
the polymer and that the zones rich in plasticizer are alternated with the
other ones  with a low content of plasticizer. Furthermore such arrangement
is quasiperiodic with a period close to 4 nm. This inhomogeneity of the
plasticized polymer was assumed to be due to the cohesion inhomogeneity of
the glassy polymer without plasticizer: "More cohesive are separated by
softer interdomain zones" \cite{Duv90,Mer96}. The other interesting
experimental result is the increase of the excess of low-frequency Raman
scattering (LFRS) or Raman boson peak by plasticization: the boson peak is
more intense than the one obtained by adding the LFRS of the two components
(PMMA and DBP) according to the composition of the plasticized material. In
the frame of the considered interpretation, the intensity of the boson peak
is related to the inhomogeneous cohesion: the more contrasted the cohesion
at the nanoscale, the more intense the boson peak. The increase of the
boson peak is interpreted by the increase of the elastic constant contrast
due to the high content of plasticizer in between cohesive domains.

The intensity of LFRS is proportional to the vibrational density of states
(VDOS) and to the light-vibration coupling coefficient. The increase of the
Raman boson peak may be due to the increase of either  the VDOS, the
coupling coefficient or both. The excess of VDOS can be observed by
inelastic neutron scattering  and will be called neutron boson peak. It is
why low frequency inelastic neutron scattering (LINS) measurements were
carried out on plasticized and non-plasticized PMMA. On the other hand, in
order to know the effect of vibrations of one component on the other, two
different plasticized PMMA were compared: the one with hydrogenated PMMA,
and the other with deuterated PMMA. Because the incoherent inelastic
neutron scattering is much higher for the hydrogenated polymer than for the
deuterated one, it will be possible to know the contribution of each
component in plasticized PMMA. In this paper the VDOS of different samples
will be compared: pure hydrogenated PMMA, pure DBP, plasticized
hydrogenated and deuterated PMMA. It will be shown that the plasticization
has no effect on the VDOS excess or neutron boson peak, and on the other
hand that the low-frequency vibrations of PMMA drag along the motion of
the DBP plasticizer.

\section{EXPERIMENT}

\subsection{Samples}

Plasticized or non-plasticized film samples, in the form of the circles of
5 cm diameter and ca. 0.3 mm in thickness, were prepared by casting the
solutions of PMMA or  PMMA/DBP blends  in
toluene/dioxane/dichloroetane/acetone mixture onto glass surface. Removing
the solvents was carried out under slow evaporation conditions for 24 h at
room temperature, with subsequent step-like heating at 80, 100 and
120$^{0}C$, up to the constant weight of the films. Four samples were
compared: pure hydrogenated PMMA (PMMA-H), pure dibutyl phtalate (DBP),
hydrogenated PMMA with 23 (mass) $\%$ DBP (PMMA-H/DBP), deuterated PMMA
with 23 (mass) $\%$ DBP (PMMA-D/DBP). The respective glass transition
temperatures, as estimated by differential scanning calorimetry (heating
rate of 20 K/min), are
$390\,K$ for pure PMMA sample $190\,K$ for DBP, and $320\,K$ for the
plasticized samples. The number average molecular weight of PMMA-H and
PMMA-D is approximately equal to 500 000 g/mol. The percentage of
deuteration of PMMA-D is $98\%$.

\subsection{Neutron scattering}

The inelastic neutron spectra were recorded on the time-of-flight
instrument IN6 at the ILL, Grenoble. The wavelength of the incident
neutrons was equal to $5.12\,\AA$ resulting in an elastic resolution (FWHM)
of $80\,\mu eV$, and an elastic momentum transfer range extending from
$Q=0.22\,\AA^{-1}$ to $Q=2.06\,\AA^{-1}$. The spectra were recorded at
three different temperatures, at $4\,K$ for the determination of
resolution, $30\,K$ and $300\,K$, using a helium cryofurnace. The
temperature of $30\,K$ was chosen to obtain the neutron inelastic
scattering by the low-frequency harmonic vibrations without the scattering
by anharmonic or relaxational motions negligible at this temperature
\cite{Mer96}. The scattering cross-sections were obtained after the usual
standard calibrations by means of the vanadium runs and the removal of the
empty cans contributions. The VDOS for harmonic modes were obtained by
taking the average of the spectra given by the different detectors, i.e.,
the average over the range from $Q=0.22\,\AA^{-1}$ to $Q=2.06\,\AA^{-1}$.
Because the first sharp peak of PMMA in the static structure factor is at
$Q=0.95\,\AA^{-1}$, a value that is lower than the upper limit of the
experimental Q-range, the incoherent approximation was applied by using the
total neutron scattering bound cross-section ($\it{coherent+incoherent}$).
The VDOS were calculated  through the use of an iterative procedure
described elsewhere \cite{Fon90}. The so-obtained VDOS were corrected by
the Debye-Waller factor and for the multiphonon contributions.\par

\section{EXPERIMENTAL RESULTS}

The VDOS divided by the square of energy, $G(E)/E^{2}$, are plotted in
Figure 1. This type of plot is conventional for comparison with the Debye
regime. It was not possible to obtain the absolute VDOS. However, as  the
shapes of the $G(E)/E^{2}$ curves for the different samples were observed
to be identical at the energies higher than 4 meV, a normalization was
obtained by coincidence of the curves from an energy of 4 meV. Obviously
this normalization does not allow to compare the total VDOS of the
different samples, but it makes possible the comparison of the boson peaks,
which appear around 2 meV.

The boson peak appears at an energy slightly lower than 2 meV for pure
PMMA-H, as observed before \cite{Mer96}, and higher for pure DBP (Figure
1). The relative intensity  at low energy is higher for pure PMMA than for
pure DBP, and decreases with plasticization. However, although the LINS of
PMMA-D is expected to be very low, in view of its relatively weak neutron
bound cross-section \cite{Be88},$G(E)/E^{2}$ of PMMA-D/DBP and of
PMMA-H/DBP are not very different. On the other hand, it is noted that the
slope of $G(E)$ or $G(E)/E^{2}$, on the low-energy side of the boson peak
is the steepest for the pure DBP.

\section{INTERPRETATION}

\subsection{Model}
From a simple glance of the curves in Figure 1, one expects that the
plasticization does not increase the boson peak relatively to the one
obtained from the addition of the VDOS of respectively the PMMA and DBP
components. It is confirmed by Figure 2. As it can be observed,
$G(E)/E^{2}$ of PMMA-H/DBP is very well fitted by adding the VDOS of
respectively PMMA-H and DBP in the  following proportion:

\begin{eqnarray}
\label{plast1}
G_{\text{PMMA-H/DBP}}(E)=aG_{\text{PMMA-H}}(E)+bG_{\text{DBP}}(E)
\end{eqnarray}

\noindent
The  low-energy VDOS of the plasticized glassy polymer obeys the general
addition law:

\begin{eqnarray}
\label{plast2}
 g(E)=\sum_{i}C_{i}g_{i}(E)
\end{eqnarray}

\noindent
where $g(E)$ and $g_{i}(E)$ are the absolute VDOS, and the coefficient
$C_{i}$  the (mass) concentration of the component $i$. In our PMMA-H/DBP,
for instance, $C_{i}$ is 0.77 and 0.23 for respectively PMMA-H and DBP. The
coefficients a and b in \ref{plast1} are  proportional to 0.77 and 0.23
respectively. The good fit in Figure 2 demonstrates that the plasticization
does not enhance the neutron boson peak or the low-energy VDOS like it does
for the Raman boson peak \cite{Duv98}.

In order to try to fit in a similar manner $G(E)/E^{2}$ of PMMA-D/DBP, it
was taken into account that the observed  VDOS, $ G_{i}(E)$, depends on the
total neutron bound cross-sections of the different atoms in the molecule.
For the low-energy vibrational modes, which are mainly dependent on the
intermolecular bonding and are studied in this work, the neutron bound
cross-section proportionality of $ G_{i}(E)$ is approximately the following:

\begin{eqnarray}
\label{plast3}
 G_{i}(E) \propto \beta_{i}g_{i}(E)
\end{eqnarray}

\noindent
where:

\begin{eqnarray}
\label{plast4}
\beta_{i}=\sum_{j} c_{j} \sigma_{j}/M_{j}
\end{eqnarray}

\noindent
with $c_{j}$, $\sigma_{j}$ and $M_{j}$ respectively the (mass)
concentration, the total neutron bound cross-section and the  mass of atom
$j$ in the molecule or monomer $i$.

From these equations, the fit of $G(E)/E^{2}$ for PMMA-D/DBP corresponding
to that for PMMA-H/DBP (\ref{plast1}) is given by the following equation:

\begin{eqnarray}
\label{plast5}
G_{\text{PMMA-D/DBP}}(E)=a(0.98\rho+0.02)G_{\text{PMMA-H}}(E)+bG_{\text{DBP}}(E)
\end{eqnarray}

\noindent
In this equation the fractions 0.98 and 0.02 accounts for the degree of
PMMA deuteration , that is equal to 0.98. The coefficient $\rho$ is equal
to $\beta_{\text{PMMA-D}}/\beta_{\text{PMMA-H}}$. Using the total neutron
bound cross-sections of the different atoms in PMMA-D and PMMA-H given in
\cite{Be88}, it is found $\rho=0.079$. This small value is due to the very
large value of the incoherent cross-section of hydrogen in comparison with
the total cross-sections of the other atoms including  deuterium. The
$G(E)/E^{2}$ curve given by (\ref{plast5}) is plotted in Figure 2. As
expected, due to the weak value of $\rho$, this calculated $G(E)/E^{2}$
curve is not very different from the experimental one for pure  DBP. One
observes in Figure 2, that the experimental $G(E)/E^{2}$ of PMMA-D/DBP is
more intense at low energies than the corresponding one calculated by
(\ref{plast5}). Furthermore it is remarked (Figure 1) that the $G(E)/E^{2}$
shapes of PMMA-H/DBP and PMMA-D/DBP are similar. The difference between the
experimental and calculated $G(E)/E^{2}$ of PMMA-D/DBP is explained by the
motion of the DBP plasticizer dragged along by the low-energy vibrations of
PMMA-D and reciprocally.  This difference is so large that it is impossible
to explain it by an underestimation  of the PMMA-D VDOS in that of
PMMA-D/DBP.

\subsection{Discussion}

The first interesting result is the good fit of  PMMA-H/DBP $G(E)/E^{2}$
obtained by the addition of the PMMA-H and DBP VDOS (\ref{plast1}). It
means that the neutron boson peak is not enhanced by the plasticization:
the low-energy VDOS of the plasticized sample is equal to the sum of the
component ones. As noted in the previous subsection, the low-energy VDOS
depends on the intermolecular bonding. In consequence, if  the VDOS of the
plasticized PMMA is given by the sum of those of PMMA and DBP, one deduces
that the bonding, on the one hand,  between PMMA macromolecules, and on the
other hand, between DBP molecules, is not significantly changed by
plasticization. As the low-energy modes observed in the boson peak are
extended on nanometric distances, such unchanged intermolecular bonding is
possible if, at the nanoscale, there is separation of both components. This
interpretation confirms the observation by SAXS of a correlation  peak at
$1.5\, nm^{-1}$ \cite{Duv98}. The distribution of DBP in PMMA is
quasiperiodic.

The non-enhancement of the neutron boson peak or of the low-energy VDOS,
means that observed enhancement of the Raman boson peak by plasticization
is due to the amplification of the vibration modulated polarisability. This
is explained by the increase of the spatially fluctuating static
polarisability due to the heterogeneous  distribution of the plasticizer in
PMMA and the nanodomain vibrations, which have a strong amplitude at the
interface between  PMMA and DBP phases. This explanation is consistent with
the interpretation of the boson peak due to an inhomogeneous cohesion of
the glassy network \cite{Duv90,Mer96}.

The second interesting result is the excess of the experimental
$G(E)/E^{2}$ of PMMA-D/DBP at low-energies in comparison with $G(E)/E^{2}$
calculated by using the total neutron bound cross-sections of atoms in
PMMA-D. Because the VDOS of the plasticized PMMA is given by the sum of the
components, as shown above, one deduces, while the total VDOS is conserved,
there is a redistribution of the low-energy VDOS among the PMMA and DBP
components in the plasticized glassy polymer. As noted above the vibrations
of one component drag along the motion of the other. This VDOS
redistribution is more effective for the vibrational modes on the
low-energy side of the boson peak (Figure 2). The reciprocal partition of
vibrational amplitude of both components will be more effective for the
vibrations which have a strong amplitude at the interface between PMMA and
DBP phases. Again this explanation is in agreement with the model, in which
the modes on the low-energy side of the boson peak correspond to the
fundamental modes of the nanodomains \cite{Duv90,Mer96}.

\section{CONCLUSION}

The total low-energy vibrational density of states of PMMA plasticized by
DBP is well fitted by the sum of the component ones. It means that the
plasticization does not change significantly the PMMA and DBP
intermolecular bonding. However, both components do not vibrate
independently: vibrations of PMMA, appearing in the low-energy side of the
boson peak, were observed in the inelastic neutron scattering of DBP. This
behavior means that the neutron boson peak is not enhanced by
plasticization. Therefore, the enhancement by plastcization of the boson
peak observed by Raman scattering comes from the  increase of the  electric
polarizability modulated by low-energy vibrations , and not from the
increase of the vibrational density of states. These results are in
agreement with an inhomogeneous distribution of the plasticizer at the
nanoscale, as it was observed by small angle X-ray scattering, and, on the
other hand, with an inhomogeneous intermolecular bonding in polymeric
glasses.

\newpage

\begin{figure}
\label{f1}
\caption{Vibrational density of states divided the square of energy,
$G(E)/E^{2}$, deduced from the inelastic neutron scattering at $30\,K$.
PMMA: empty circles; PMMA-H/DBP: empty squares; PMMA-D/DBP: full squares;
DBP: full circles The normalization is obtained by an arbitrary
equalization of $G(E)/E^{2}$ at 4 meV. }
\end{figure}

\begin{figure}
\label{f2}
\caption{Comparison of the measured vibrational density of states with the
calculated ones. Density of states of plasticized hydrogenated (empty
squares) and deuterated (full squares) poly(methyl methacrylate) compared
with the ones calculated by Equations \ref{plast1} (full line) and
\ref{plast5} (dashed line) respectively.}
\end{figure}


\begin{references}

\bibitem{Duv98}E. Duval, M. Kozanecki, L. Saviot, L. David, S. Etienne, V.
A. Bershtein and V. A. Ryzhov, Europhys. Lett. {\bf 44}, 747 (1998)

\bibitem{Duv90}E. Duval, A. Boukenter and T. Achibat, J. Phys. Condens.
Matter {\bf 2}, 10227 (1990)

\bibitem{Mer96}A. Mermet, N. V. Surovtsev, E. Duval, J. F. Jal, J.
Dupuy-Philon and A. J. Dianoux, Europhys. Lett. {\bf 36}, 277 (1996)

\bibitem{Fon90}A. Fontana, F. Rocca, M. Fontana, B. Rossi and A. J.
Dianoux, Phys. Rev. B {\bf 41}, 3778 (1990)

\bibitem{Be88}M. B\'{e}e, in {\it Quasielastic neutron scattering}, (A.
Hilger, Bristol, 1988)

\end{references}
\end{document}